\documentclass[a4paper, amsfonts, amssymb, amsmath, reprint, showkeys, nofootinbib, twoside, superscriptaddress,longbibliography]{revtex4-2}
\usepackage[english]{babel}
\usepackage[utf8]{inputenc}
\usepackage{gensymb}
\usepackage[colorinlistoftodos, color=green!40, prependcaption]{todonotes}

\usepackage{siunitx}
\usepackage{lipsum}

\usepackage[pdftex, pdftitle={Article}, pdfauthor={Author}]{hyperref} 
\begin{document}
\title{A story with twists and turns : how to control the rotation of the notched stick}
\author{Martin Luttmann}
    \email[Correspondence email address: ]{martin.luttmann@gmail.com}
    \affiliation{Universit\'e Paris-Saclay, CEA, LIDYL, 91191 Gif-sur-Yvette, France}
    \affiliation{DQML, IMX,
École polytechnique fédérale de Lausanne (EPFL) Station 12,
CH-1015 Lausanne, Switzerland}

\author{Michel Luttmann}
    \email[Correspondence email address: ]{michel.luttmann@cea.fr}
    \affiliation{CEA, CESTA, F-33114 Le Barp, France}

\date{\today} 

\begin{abstract}
The notched stick, also known as the Gee-Haw-Whammy-Diddle, is a wooden toy able to convert linear vibration into rotational motion, whose behavior has been intriguing both children and physicists for decades. The oldest scientific article one can find on this subject was published 87 years ago in the present journal. Here we derive an analytical model of the system, supported by experimental results. We predict the direction of rotation, and explain why the device is so easy to operate, even without fine control of the various parameters. 
The potential importance of the vertical displacement of the finger exerting the perturbation force is also highlighted. 
 We finally discuss similarities between the mechanical system described here and the optical effect of birefringence.   


\end{abstract}


\maketitle

\section{Introduction}
\begin{figure}
    \centering
    \includegraphics[width=\linewidth]{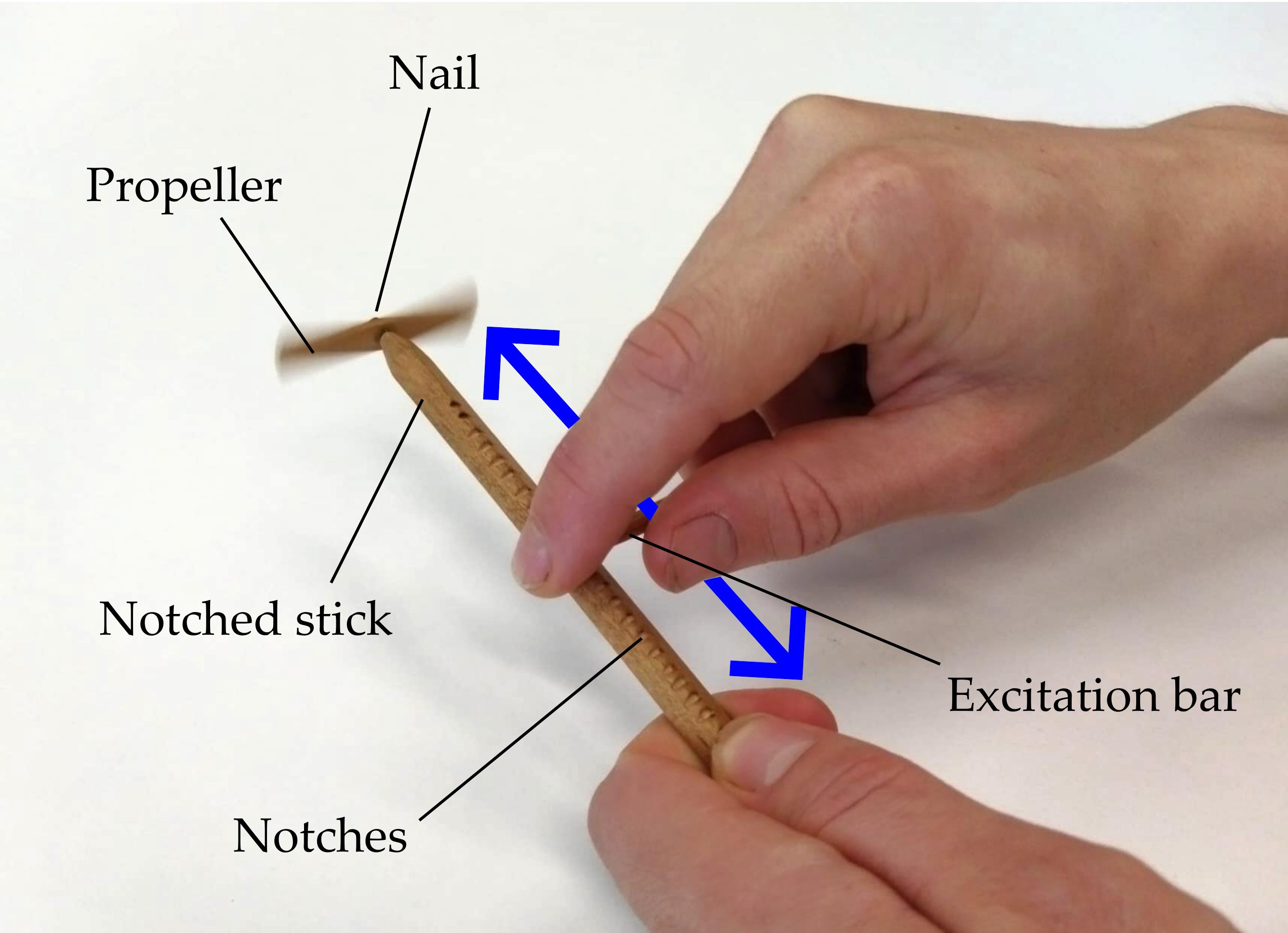}
    \caption{(color online) A typical configuration of the device. Here, the notched bar has a square cross-section. The excitation bar is used to rub the notched stick back and forth (blue arrow), inducing a vertical vibration of the nail on which the propeller is mounted.}
    \label{fig:photo}
    \end{figure}

What physicist has not marveled, as a child, at a beautiful demonstration involving, for instance, soap bubbles, tippe tops, or Newton's pendulum? The so-called \textit{notched stick} is one of those scientific toys that stimulate the curiosity of children \cite{Scarnati1992} and adults alike. It is an old wooden toy, made of a rod with a nail at the end. The nail serves as a shaft for a small propeller. A series of transverse notches is engraved along a portion of one edge of the stick. The player holds the notched stick in one hand, and a small cylindrical wooden dowel in the other. This second bar is used to rub the stick along its notched edge (Fig.\,\ref{fig:photo}), causing it to vibrate and emit a sound close to that of a rattle. Experienced players are able, using a method described below, to convert this vibration into a rotation of the propeller. As we shall see, the direction of rotation can be controlled at will by the player. The origin of this toy is sometimes attributed to Native Americans, while it is known today by different names around the world: Gee-Haw-Whammy-Diddle \cite{Aubrecht1982}, Hooey stick \cite{Scarnati1992} or Ouija Windmill in North America, Hui-Maschine \cite{Schlichting1988} in Germany, Girigiri-Garigari \cite{Satonobu1995} in Japan, etc. In France, the player pronounces the word \textit{Bozo-Bozo} \cite{JCBone1994, Courty2004} as a magic spell before the sudden reversal of the rotation, for the greatest delight of his/her audience. 

\begin{figure}
    \centering
    \includegraphics[width=0.85\linewidth]{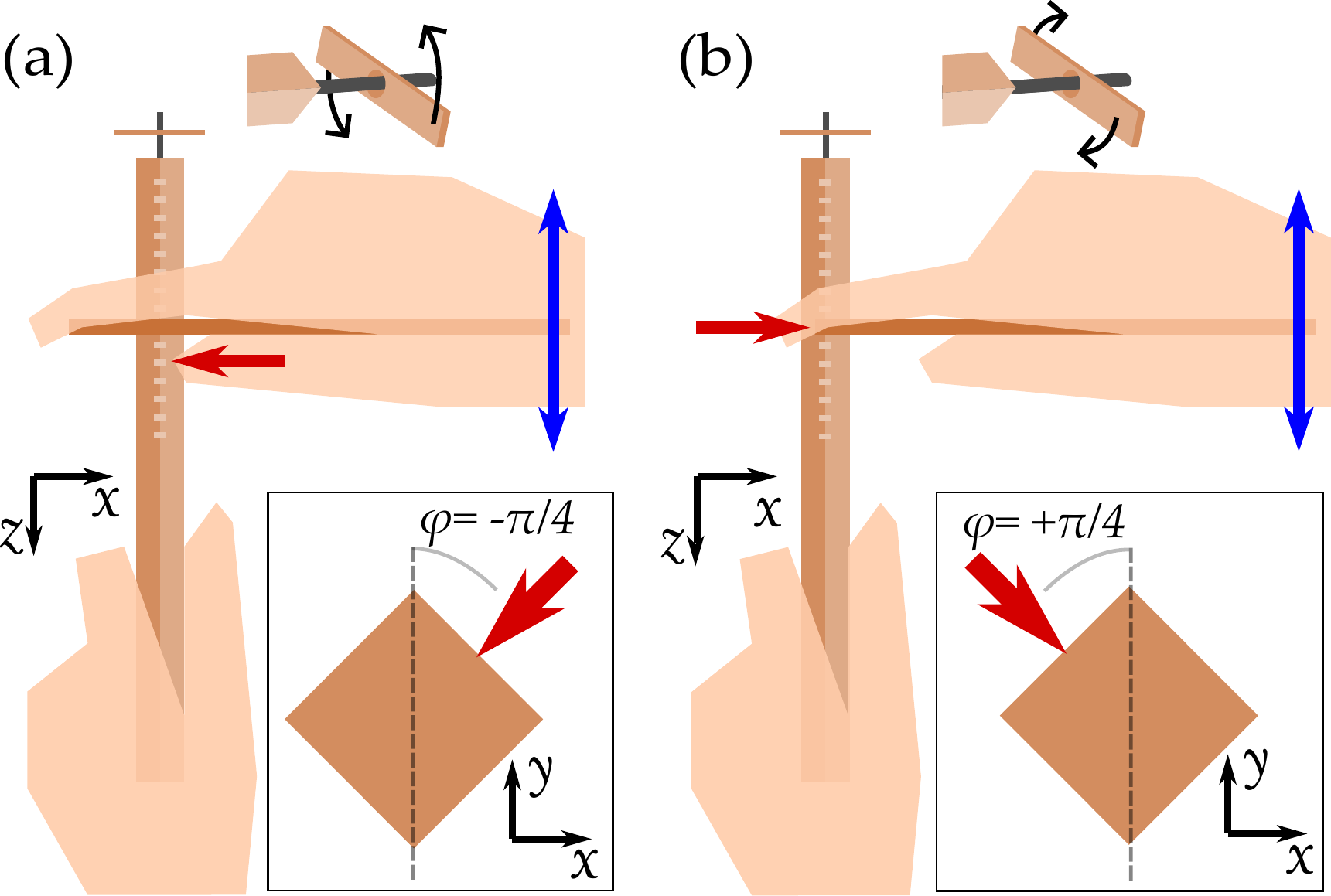}
    \caption{(color online) (a) In the case where the thumb of the hand holding the excitation bar is used to push on the top right face of the notched stick (red arrow, and inset image), the propeller starts spinning counter-clockwise from the point of view of the user. (b)  If then the force is exerted on the top left face of the notched stick (red arrow, and inset image), the propeller spins in the opposite direction. Here $\varphi$ denotes the angle of the perturbation force with respect to the vertical axis.}
    \label{fig:fig1}
    \end{figure}

In the following, we will refer to the bar with notches as the \textit{notched stick}, and to the additional stick used to rub the notches as the \textit{excitation bar}.
To control the propeller's direction of rotation, the player uses the following trick. As they rubs the notched stick with the excitation bar, they brings one of the fingers of the hand holding the latter into contact with the notched stick \cite{Welch1973, Aubrecht1982, Scarnati1992}. For a right-handed player, if the thumb is gently pushed against the top right-hand face of the stick, the propeller is driven in a counter-clockwise motion from their perspective (Fig.\,\ref{fig:fig1}.a). Conversely, if the tip of the index finger is pressed on the top-left face of the notched stick, the rotation is clockwise (Fig.\,\ref{fig:fig1}.b). By discreetly pressing on one side and then the other, the player thus reverses the direction of rotation (Supplementary video 1). If the player does not press on either side, the rotation is random, or does not take place at all.

A notched stick can be built in minutes, without worrying about the exact dimensions of the stick or the spacing and depth of the notches. With a bit of practice, the device is then surprisingly easy to operate. However, its working principle has been intriguing physicists for almost a century \cite{Leonard1937}, and a dozen of scientific studies have been reported so far. Most of the papers published on the subject aim at describing the coupling between the motions of the nail and that of the propeller, assuming linear or elliptical \cite{Wilson1998} vibration of the nail. Recently, Marek \textit{et al.} \cite{Marek2018} and Broseghini \textit{et al.} \cite{Broseghini2019} demonstrated experimentally that a purely linear oscillation of the nail can lead to a rotation of the propeller, amplified by ordinary and/or parametric resonance (a phenomenon resembling the hula-hoop effect \cite{Caughey1960, Cross2021}, or Kapitza pendulums \cite{Bhattacharjee2013}). This is a case of spontaneous symmetry breaking, caused by the mismatch between the position of the nail and the propeller's center of mass. Although there seem to be no definite consensus on the subject, these studies offer two major lessons. First, in almost all cases, a rotation of the pivot point (the nail) leads to a rotation of the propeller in the same direction, just as we do intuitively when twirling a hula-hoop around the hips, or a ring around the arm. Second, a linear oscillation of the pivot can also trigger a rotation of the propeller, although the direction is then seemingly random and determined by fluctuations or by the exact initial conditions \cite{Bhattacharjee2013,Broseghini2019}. Hence, the mechanism of control of the propeller's rotational direction is to be found in the origin of the motion of the nail itself.

A small number of papers have discussed the mechanism of control of the trajectory of the tip of the notched stick \cite{Leonard1937, Miller1955, Laird1955, Scott1956, Welch1973, Aubrecht1982, Schlichting1988, Satonobu1995}, often not including a quantitative analysis. The most comprehensive studies of this aspect have been performed by Leonard \cite{Leonard1937} and Satonobu \textit{et al.} \cite{Satonobu1995}. In his theoretical work, Leonard considered a different configuration of the device, where the notched stick's cross section is rectangular and the excitation bar positioned diagonally, and attributed the working mechanism to a difference of resonance frequency between the two cartesian components of the motion. This picture has several limitations. First, later studies \cite{Satonobu1995} showed that the bending of the notched stick is negligible in this problem. Second, we observe that notched sticks with square or circular cross sections are also operational. Finally, the author does not precisely predict the direction of rotation or explains the mechanism for its reversal.
Although they could not make quantitative predictions, Satonobu \textit{et al.} performed valuable experiments and proposed instead that a difference between the waveforms of the force applied in the direction of the perturbation finger and the direction perpendicular to it could explain the behavior of the toy. We will discuss this idea in Section \ref{sec: discussion} of this article. This intuition, however, lacks the support of a rigorous model that provides testable predictions.

In this paper, we clarify and unify previous approaches in a simple, more general mechanical model in order to investigate the origin of the rotation of the nail supporting the propeller. The end of the notched stick is modelled as a point mass moving in a 2D plane under the influence of various forces. Crucially, the point mass dynamics is influenced by two restoring forces corresponding to the action of the hand holding the notched stick and that of the finger pushing on it with adjustable direction \cite{Vsauce}. Contrary to previous approaches, our model successfully predicts the direction of rotation, and highlights the role of the energy loss, which, in particular, determines the robustness of the mechanism. The model also accounts for the fact that the only parameter determining the direction of rotation is the pushing angle of the perturbation finger. New behaviors are also predicted, such as the dependence of the trajectory on the driving frequency. Next, we describe an experiment with an idealized version of the notched stick, driving its motion with a sinusoidal force and show that the experimental results are consistent with our model. Finally, we propose that the notched stick undergoes an additional effect, linked to the vertical displacement of the perturbation finger, which may lead to a more pronounced rotation. In the following, we will refer to the device shown in Fig.\,\ref{fig:photo} as the \textit{real-life notched stick}, to distinguish it from our laboratory version of the device. We also refer to the finger controlling the rotation as the \textit{perturbation finger}. 

\section{Analytical model} 

\begin{figure}
    \centering
    \includegraphics[width=\linewidth]{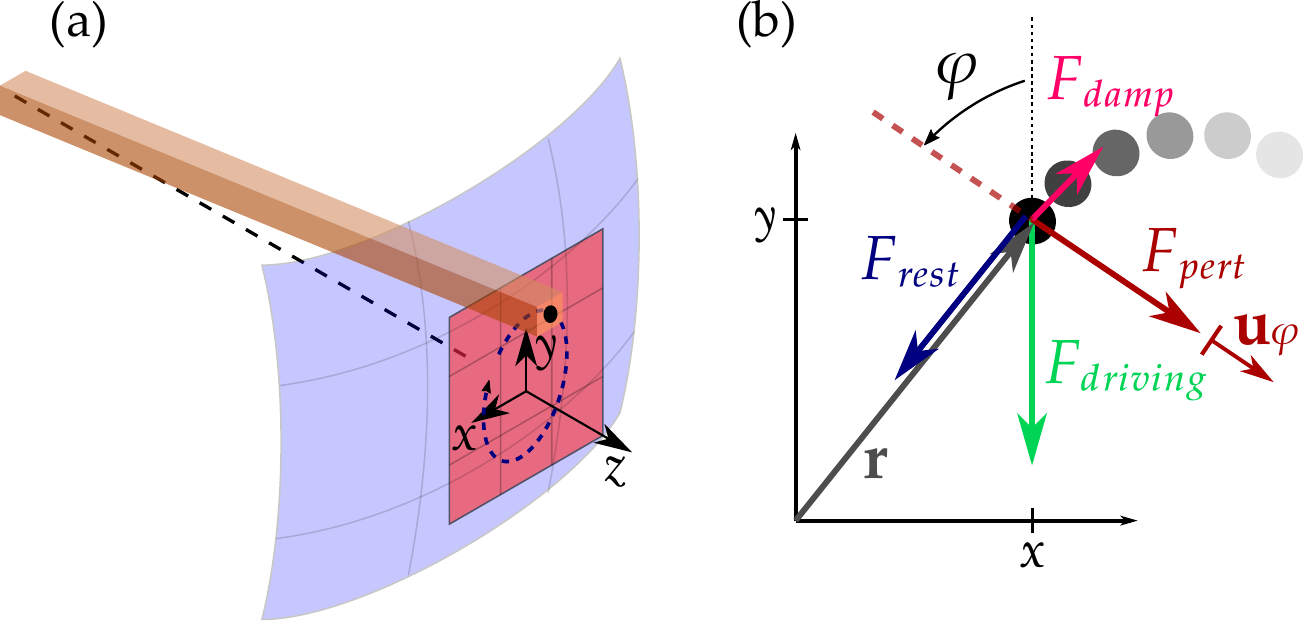}
    \caption{(color online) (a) For small oscillations of the notched stick around the $z$ axis, the trajectory of its tip is confined in a vertical plane (shown in red). (b) The analog point mass (black dot) is acted on by four distinct forces in the $(x,y)$ plane.}
    \label{fig:fig2}
    \end{figure}

As previous studies demonstrated that the bending of the notched stick is negligible \cite{Satonobu1995}, we consider a perfectly rigid notched stick, initially parallel to the $z$ axis. We assume that one end of the stick is fixed in space, while the other is free to move. Additionally, since the amplitude of the motion of the real-life notched stick's tip is tiny compared to its length, we assume that the notched stick angle with respect to the $z$ axis remains small. Therefore, the tip of the stick is approximately confined in the $(x,y)$ plane (Fig.\,\ref{fig:fig2}.a), and its position vector in this plane is denoted $\textbf{r}$. 

In this model, the oscillation of the notched stick's tip is fully analogous to that of a point mass moving in a 2D plane under the influence of four distinct forces (Fig.\,\ref{fig:fig2}.b). A driving force $\textbf{F}_\text{driving}(t)$ is directed along the $y$ axis, and corresponds to the periodic force exerted by the excitation bar as it rubs the notches. A restoring force $\textbf{F}_\text{rest}(\textbf{r})$ pulls the point mass back to the origin, mimicking the effect of the flexible hand holding the real-life notched stick. We assume $\textbf{F}_\text{rest}(\textbf{r}) = - k \textbf{r}$, where $k$ is the stiffness constant of the holding hand. In order to take into account the effect of the perturbation finger, an additional force $\textbf{F}_\text{pert}$ is applied on the stick at an angle $\varphi$ away from the $y$ axis (for instance Fig.\,\ref{fig:fig1}.a corresponds to $\varphi = -\pi/4$). To take into account the flexibility of the perturbation finger, $\textbf{F}_\text{pert}$ is expressed as an elastic force acting along the $\textbf{u}_\varphi$ vector (see Fig.\,\ref{fig:fig2}.b)
\begin{equation}
    \textbf{F}_\text{pert}(\textbf{r}) = - q (\textbf{r} \cdot \textbf{u}_\varphi ) \textbf{u}_\varphi,
    \label{eq: fpert}
\end{equation}
where $q$ is the stiffness constant of the perturbation finger. This additional force breaks the mirror symmetry of the system, thus allowing for clockwise or counter-clockwise oscillations of the stick. From an energy perspective, the combined forces $\textbf{F}_\text{rest}$ and $\textbf{F}_\text{pert}$ correspond to an anisotropic energy potential, as mentioned in Ref.\,\cite{Scott1956} which compares the notched stick to a ball rolling in a distorted watch glass. Finally, a damping force, that we write for simplicity $\textbf{F}_\text{damp} = -\alpha \frac{\text{d}\textbf{r}}{\text{d}t}$, accounts for the energy loss.
The point mass then obeys Newton's second law
\begin{equation}
    m \frac{\text{d}^2\textbf{r}}{\text{d}t^2} = \textbf{F}_\text{driving} + \textbf{F}_\text{rest}
    + \textbf{F}_\text{pert}
    +\textbf{F}_\text{damp},
    \label{eq:general}
\end{equation}
where $m$ is the equivalent  mass associated to the notched stick's. 
Let us now consider the simplest possible driving force, a vertical sinusoidal force of frequency $\omega$: $\textbf{F}_\text{driving}(t) = F_0 e^{i \omega t} \, \textbf{u}_y$, where $\textbf{u}_y$ is the unit vector along $y$. Note that the back and forth motion of the excitation bar actually creates a variable torque on the notched stick, resulting in an additional low frequency component of the driving force. This additional component is neglected here for simplicity. 
To compute the steady-state solution, we express the cartesian coordinates of the point mass as $x(t) = x_0 \, e^{i \omega t}$ and $y(t) = y_0 \,e^{i \omega t}$, where $x_0$ and $y_0$ are complex numbers. Plugging these and the expression of the four forces in Eq.\,\ref{eq:general}, we obtain
\begin{equation}
\begin{split}
m\omega^2 x(t) &= \\ 
 k x(t) &+q \big( x(t) \sin^2 \varphi - y(t) \cos \varphi \sin \varphi \big) + i \omega \alpha x(t) \\
m\omega^2 y(t) &= \\
k y(t) &+q \big(y(t) \cos^2 \varphi -x(t) \cos \varphi \sin \varphi\big) + i \omega \alpha y(t) \\
&+ F_0e^{i \omega t} .
\end{split}\label{eq:system}
\end{equation}
Dropping the $t$ dependence, we rewrite the first of the two above equations as
\begin{equation}
    x = \xi \cdot y,
    \label{eq: x del y}
\end{equation}
with
\begin{equation}
    \xi = \frac{q \cos \varphi \sin \varphi}{m (\omega_0^2 -\omega^2)  +i \alpha \omega},
    \label{eq:Delta}
\end{equation}
and where
\begin{equation}
    \omega_0 = \sqrt{(k+q\sin^2\varphi)/m}.
    \label{eq: w0}
\end{equation}
The modulus of $\xi$ corresponds to the ratio of the amplitudes of the $x$  and $y$ components of the motion, while its complex argument provides us with the dephasing between them. Since the argument of $\xi$ is not zero in general, the trajectory of the notched stick's tip is generally an ellipse. Eq.\,\ref{eq: x del y} alone is sufficient to compute the shape of the trajectory, while the full solution of Eq.\,\ref{eq:general} is obtained by plugging Eq.\,\ref{eq: x del y} in the second row of Eq.\,\ref{eq:system} (Supplementary Materials S1).

First, we note in Eq.\,\ref{eq:Delta} that a $\pi/2$ variation of the perturbation angle $\varphi$ changes $\xi$ to $-\xi$, which corresponds to a $\pi$ phase shift between the two cartesian components of the motion. It thus yields a reversal of the ellipse’s direction of travel. Hence, the angle $\varphi$ determines the direction of rotation of the stick’s tip, as observed with the real-life notched stick. For instance, $\varphi = \pi/4$ corresponds to counter-clockwise rotation, in agreement with the practical observation.

Secondly, we note in Eq.\,\ref{eq:Delta} that values of $\varphi$ that are integer multiples of $\pi/2$ yield $\xi = 0$. In that case, we have $x(t)=0$, thus the motion is linear vertical. On the other hand, if $\varphi$ is an odd multiple of $\pi/4$, the modulus of $\xi$ is maximized, allowing non-degenerate elliptical motion. Note that the square cross-section of the notched stick (Fig.\,\ref{fig:fig1}b,c) results in a non-zero component of the perturbation force pointing in the direction $\varphi = \pm \pi/4$, even if the player's finger pushes horizontally. 

Equation.\,\ref{eq:Delta} also indicates that at the resonance frequency, $\omega = \omega_0$, $\xi$ is a purely imaginary number. As such, at the resonance frequency, the components $x(t)$ and $y(t)$ are $\pm \pi/2$ out of phase, the sign of the dephasing (determining the rotation direction) being that of $\cos \varphi \sin \varphi$. The direction of rotation thus switches every $\pi/2$ variation of $\varphi$, as observed with the real-life device. If, in addition, the modulus of $\xi$ is close to $1$, then the trajectory of the notched stick's tip is close to a circle.

\begin{figure}
    \centering
    \includegraphics[width=\linewidth]{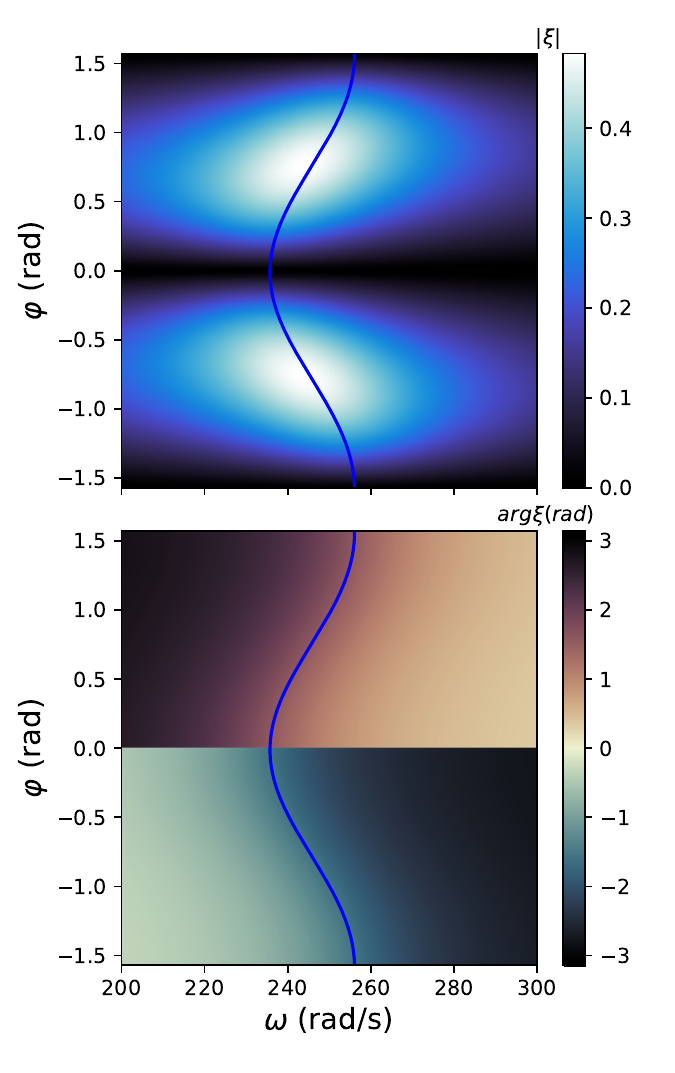}
    \caption{(color online) Modulus (top) and argument (bottom) of $\xi$, with respect to $\varphi$ and $\omega$.  Values for the parameters $k$, $q$, $m$ and $\alpha$ correspond to the experimental results described in Section \ref{sec:exp}. The blue curves show the position of the resonance frequency $\omega_0 = \sqrt{(k+q\sin^2\varphi)/m}$.}
    \label{fig:fig3}
    \end{figure}
\begin{figure*}
    \centering
    \includegraphics[width=0.9\textwidth]{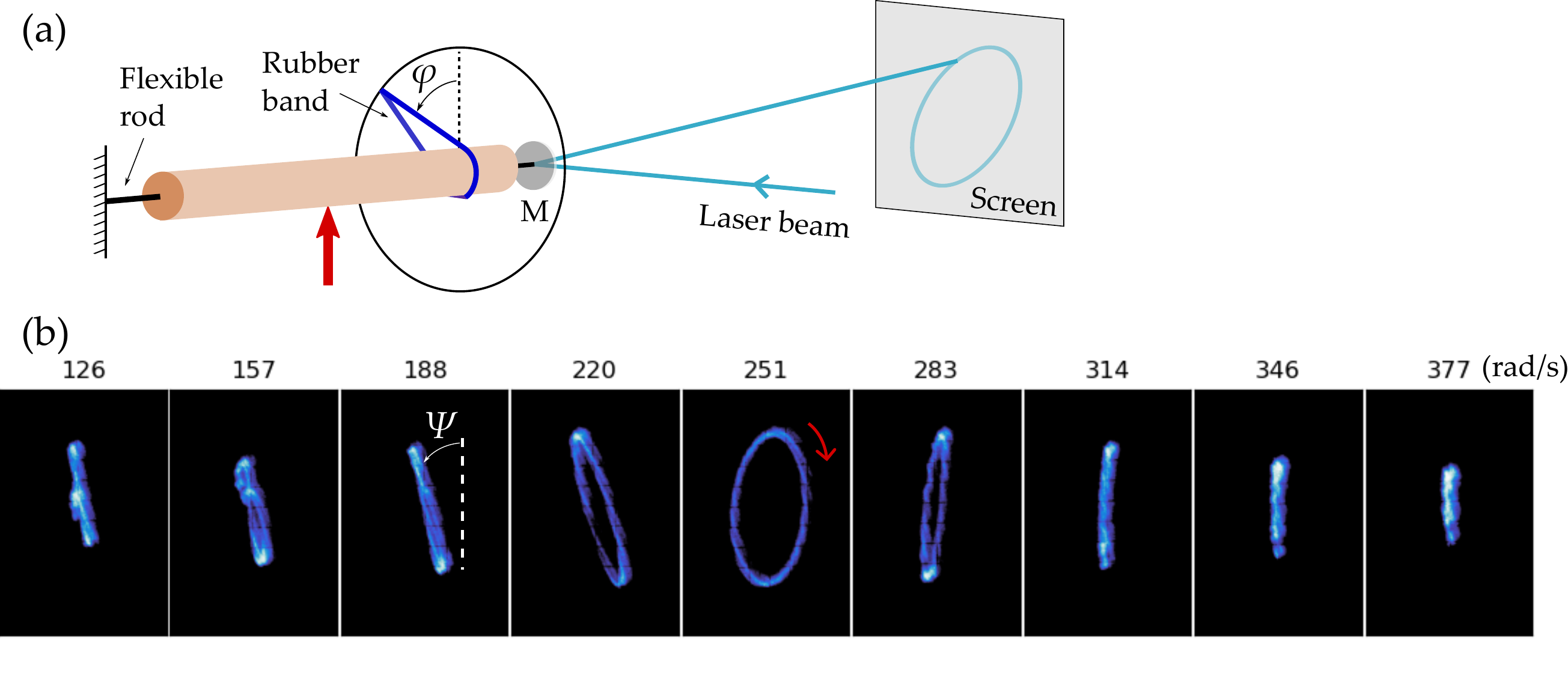}
    \caption{(color online) (a) Scheme of the experimental setup. The stick is attached with a screw to a fixed panel. A rubber band is used to pull on the stick, the angle $\varphi$ of the force applied can be varied (dark circle). A sinusoidal vertical force (red arrow) is exerted on the stick. A mirror (M) is mounted on the tip of the stick, reflecting a laser beam on a screen located 2m away from the device. (b) Experimental trajectories of the notched stick's tip for $\varphi = \pi/4$ and various values of the driving frequency. The curved red arrow shows the direction of rotation.}
    \label{fig:fig4}
    \end{figure*}
    
The aforementioned features can be observed in Fig.\,\ref{fig:fig3}, showing $|\xi|$ and the dephasing $\text{arg}(\xi)$ versus the perturbation angle and the driving frequency. The range of $\varphi$ is limited to $[-\pi/2, \pi/2]$ for clarity. The parameters $k$, $q$, $m$ and $\alpha$ of this plot correspond to the experiment reported later in the text. As expected, $|\xi|$ peaks for $\varphi = \pm \pi/4$ and $\omega$ close to $\omega_0$, while the dephasing is equal to $\pm \pi/2$ for the resonance frequency $\omega = \omega_0$ (blue curve). Note however that $|\xi|$ remains significant, and the dephasing close to $\pm \pi/2$, for a broad range of frequency centered on $\omega_0$, of width approximately given by $\alpha/m$ \cite{Landau1976}. This means that driving frequencies sufficiently close to $\omega_0$ will still excite a rotation of the device. Similarly, $|\xi|$ and the dephasing vary rather slowly with  $\varphi$. As we will discuss later in the text, this weak dependence of $\xi$ on the variables $\omega$ and $\varphi$ explains why the device continues to work even if the driving frequency and the perturbation force direction are not precisely adjusted.

\section{Experiment with an idealized version of the device} 
\label{sec:exp}
The simple model detailed above indicates that the shape of the trajectory of the notched stick's tip should depend on the driving frequency and the perturbation angle $\varphi$, while its direction of travel should only depend on $\varphi$. 

The amplitude of the oscillation of the notched stick's tip is small compared to its length, typically less than a millimeter for a 15 cm long stick. As such, recording this motion requires special techniques. Some previous experiments have used an accelerometer mounted on the tip of the notched stick, or high-speed cameras \cite{Broseghini2019, Marek2018}. Here, we propose a different, simpler method, that provides a visualization of the trajectory in real time.

The experimental setup is sketched in Fig.\,\ref{fig:fig4}.a. The idea is to magnify the oscillation using a continuous laser beam, reflected at an angle on a mirror mounted on the tip of the notched stick. The other end of the stick is screwed onto a fixed vertical panel. The flexibility of the screw allows small oscillations of the stick about its equilibrium position. The total mass of our stick is 15 grams, thus yielding an equivalent mass of $m = 5 \times 10^{-3}$ kg (this stems from the moment of inertia of a thin stick of length $l$ and mass $m$, equal to $ml^2/3$). We use a stretched rubber band to exert the perturbation force corresponding to the action of the finger in the real-life device. The direction of this force, i.e. the angle $\varphi$, can be varied by attaching the rubber band in various positions on the edge of a circular piece centered on the stick. The stiffness constant of the rubber band is estimated to be $q \approx 50$ N/m. 
A vertical sinusoidal driving force with adjustable frequency is exerted on the device by the mean of a magnetic shaker (model Ling V203). A screen is placed 2m away from the mirror in order to observe the trajectory of the laser spot. The stick being 20 cm long, the trajectory magnifying factor is about 10. Since the persistence of vision is typically longer than the period of the oscillation, one can directly see the magnified trajectory on the screen.

\begin{figure}[h!]
    \centering
    \includegraphics[width=\linewidth]{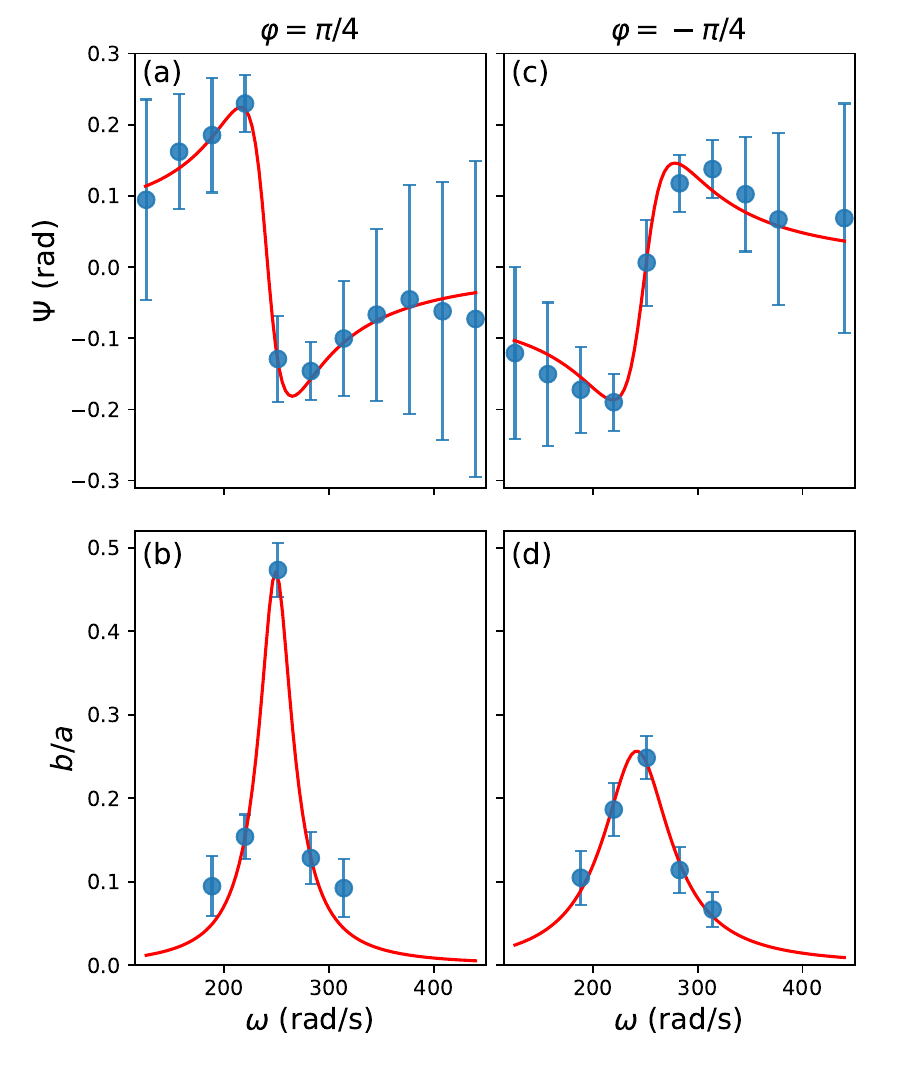}
    \caption{(color online) (a,c) Ellipse inclination angle and (b,d) ratio of the minor to the great axis, for $\varphi = \pi/4$ (left) and $-\pi/4$ (right). The blue dots are the experimental value, and the red curves are the best fits performed with our analytical model, the parameters $\omega_0$ and $\alpha$ being freely adjusted. Error bars show the parameter range for which the rms distance between the fitted ellipse and the experimental points varies by less than 10\%.}
    \label{fig:fig5}
    \end{figure}

Fig.\,\ref{fig:fig4}.b shows the trajectories photographed by a camera (0.2-s exposure time), for $\varphi = \pi/4$ and various driving frequencies $\omega$. Close to the resonance frequency, located around 250 rad/s, the trajectory takes the shape of a near-perfect ellipse. The inclination angle and eccentricity of this ellipse vary strongly with the frequency. At lower frequencies ($\omega < 200$ rad/s) and higher frequencies ($\omega > 300$ rad/s), the trajectory is almost purely linear vertical (which corresponds to a low value of $|\xi|$) with an amplitude that decreases the further we are from the resonance frequency. Fig.\,S 5 of the Supplementary Materials shows a frequency sweep performed in the same conditions for $\varphi = -\pi/4$, which exhibits a similar behavior, except for a reversal of the ellipse orientation angle and direction of rotation. These observations are in full agreement with our analytical model. However, tangled, non-elliptical trajectories are observed at certain low frequencies (for instance at $\omega = 157$ rad/s in Fig.\,\ref{fig:fig4}.b). Although we have not yet identified the origin of these imperfections, we believe they may be caused by deviations from a perfectly sinusoidal excitation force. 

To further understand the fine variations of the trajectory, we extract from these images the inclination angle $\Psi$ and the ratio $b/a$ of the minor axis to the great axis of the ellipses \cite{Collett2005}. These are shown in Fig.\,\ref{fig:fig5} for $\varphi = \pm \pi/4$, with respect to the driving frequency, and fitted with the values expected from Eq.\,\ref{eq:Delta} (Supplementary Materials S3, S4). Importantly, the resonance frequency $\omega_0$ and the damping constant $\alpha$ are adjustable parameters of the fits. Ideally, the curve of Fig.\,\ref{fig:fig5}.a should be the opposite of that of Fig.\,\ref{fig:fig5}.c, while the curves of Fig.\,\ref{fig:fig5}.b,d should be identical. We attribute the discrepancies to experimental errors in the setting of the perturbation angle $\varphi$, as well as anisotropies in the restoring force of the screw holding the notched stick to the vertical panel. Nevertheless, as the match between the experimental points and the fitted curves is satisfying, the shapes of the trajectories provide us with an estimation of the parameters $\alpha$ and $\omega_0$ in our experimental conditions. Taking the mean value of the best parameters of each fit, we find $\omega_0 = 246.7 \pm 7.3$ rad/s and $\alpha = 0.297 \pm 0.138$ kg/s (we provide here the $2\sigma$ uncertainty of the set of four values). These values were plugged in Eq.\,\ref{eq:Delta} in order to plot the graphs of Fig.\,\ref{fig:fig3}. Note that a direct measurement of $\alpha$ would have been difficult. In particular, the dynamic contact between the magnetic shaker and the stick has an influence on the energy loss of the system, hence $\alpha$ cannot be rigorously estimated by, for instance, measuring the damping time of the stick in the absence of the magnetic shaker.


\section{Discussion}
\label{sec: discussion}

The model derived here, as well as the experiment reported, considers an idealized version of the notched stick, which is quite far from the real-life device. In the case of the real-life version, the perturbation finger and the excitation bar move together along the length of the notched stick. In addition, the vertical driving force is not a sine function, but rather a pulsed excitation resulting from the periodic impacts of the excitation bar on the notches. Furthermore, the real-life notched stick is heavily damped, being held by a tight hand. For these reasons, we can wonder whether the conclusions derived above still apply to the real toy manipulated by hand.

A simple way to test the effect of the displacement of the perturbation finger is to rub the notched stick with the excitation bar, while asking another person to push laterally with his finger on a given point of the notched stick (Supplementary video 2). While it is still possible to control the direction of rotation of the propeller in this way \cite{Satonobu1995}, it is less easy than with the standard method, shown in Fig.\,\ref{fig:fig1}. 

\begin{figure}
\centering
\includegraphics[width=\linewidth]{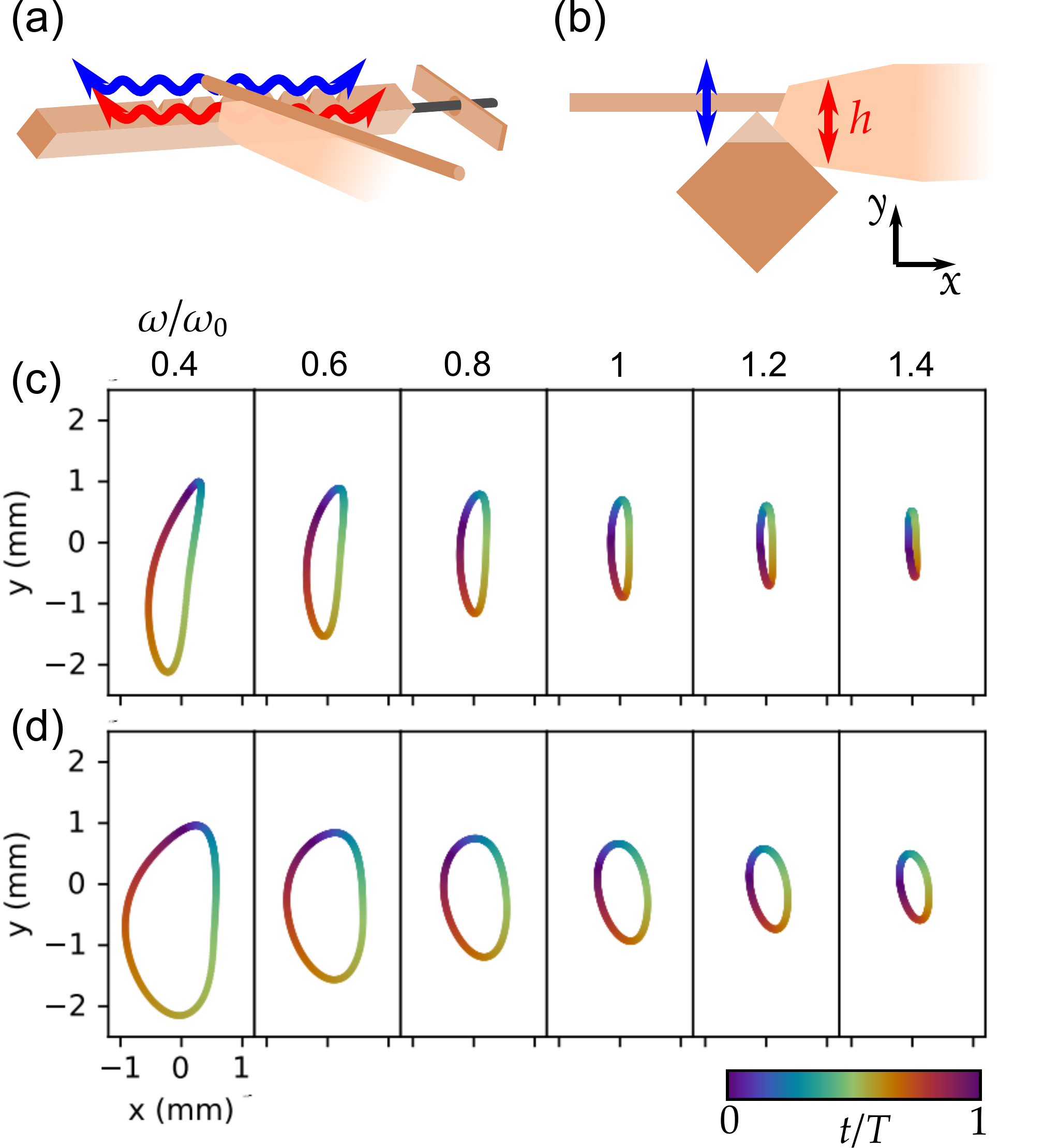}
\caption{(color online) Effect of the vertical displacement of the perturbation finger. (a) The perturbation finger is firmly connected to the excitation bar, their vertical oscillations are in synch (blue and red paths). (b) View in the $(x,y)$ plane. $h$ denotes the vertical position of the finger pushing on the notched stick. (c) Notched stick's tip trajectories for various $\omega$, for a sawtooth driving force, and $h=0$. (d) Notched stick's tip trajectories for various $\omega$, for a sawtooth driving force, and a sawtooth displacement $h(t)$ of amplitude 2 mm, dephased by $\pi/2$. Here, we have $\varphi=-\pi/4$, $q = k = 300$ N/m, and $\alpha = 2.97$ kg/s. The color codes the time corresponding to each position along the trajectory, covering a full driving period $T = 2\pi/\omega$.}
\label{fig:fig6}
\end{figure}

A possible explanation of this fact goes as follows \cite{Satonobu1995}.  The perturbation finger, pressing on either side of the notched stick, is integral with the excitation bar, and therefore moves synchronously with it (Fig.\,\ref{fig:fig6}.a). As the excitation bar oscillates vertically as it passes each notch, the perturbation finger also moves up and down periodically (Fig.\,\ref{fig:fig6}.b). The force $\textbf{F}_{\text{pert}}$ exerted by the perturbation finger must vary as a result of this vertical displacement. 
Note that this effect was not present in the laboratory device of Fig.\,\ref{fig:fig4}, since the attachment point of the rubber band exerting the perturbation force was static. To quantify the importance of this additional effect, we propose to modify the perturbation force $\textbf{F}_\text{pert}$ of Eq.\,\ref{eq: fpert} to
\begin{equation}
    \textbf{F}_\text{pert}(\textbf{r}) = - q \left((\textbf{r}- h e^{i \omega t} \textbf{u}_y) \cdot \textbf{u}_\varphi  \right) \textbf{u}_\varphi.
    \label{eq: fpert2}
\end{equation}
$h$ is a complex number whose modulus equals half the depth of the notches (about 1 mm), and whose argument corresponds to the dephasing between the vertical displacement of the finger and the driving force $\textbf{F}_\text{driving}(t)$. Supplementary Materials S5 details the computation of $x(t)$ and $y(t)$ given this new form for the perturbation force.  

In addition, we note that our analytical model allows to compute the steady-state trajectory for a single-frequency excitation. Thus, to refine the model, we compute a sum of these trajectories for different driving frequencies, corresponding to the Fourier series of a sawtooth function. Such temporal driving force is a better model of the periodic impacts of the excitation bar on the notches. In addition, we consider a more realistic situation, with a spring constant $q$ equal to $k$, which stems from the fact that the flexibility of the perturbation finger's is close to that of the holding hand. Moreover, we consider a higher value of the damping factor $\alpha = 2.97$ kg/s, equal to ten times that of our experiment. This value is greater than the critical value $\alpha = 2m\omega_0$ below which the freely moving notched stick oscillates periodically upon relaxation \cite{Feynman1964b} (which does not occur with the real-life device). 

Fig.\,\ref{fig:fig6}.c shows the trajectory of the notched stick's tip in this more realistic configuration, for $h=0$ and for various values of the driving frequency. The vertical driving force is given by the sum of the first 25 terms of the sawtooth Fourier series. Importantly, the trajectories are still close to ellipses, and are all travelled in the same (clockwise) direction. We also notice that the shape of the trajectory does not show a strong dependence over the value of $\omega$, which stems from the width of the resonance scaling as $\sim \alpha/m$. For the sake of comparison, Fig. S3 of the Supplementary Materials displays the trajectories in the same conditions, for $\alpha = 2.97 \times 10^{-1}$ kg/s, showing way more diverse trajectories. 
We conclude that the robustness of the system is conditioned by a sufficiently high value of the damping term, the real-life notched stick being heavily damped.

Secondly, in Fig.\,\ref{fig:fig6}.d, we demonstrate that a vertical oscillation of the perturbation finger ($h \neq 0$) can reinforce the rotation mechanism described in detail above (resulting in a more eccentric elliptical trajectory). These trajectories were obtained for $|h| = 2$ mm and $\text{arg}(h) = \pi/2$. We argue that close to the resonance frequency, the vertical displacement of the excitation bar is in phase with $y(t)$, but in phase quadrature with the vertical driving force $\textbf{F}_\text{driving}(t)$ \cite{Feynman1964}, effectively yielding $\text{arg}(h) = \pm \pi/2$. We therefore conclude that a vertical displacement of the perturbation finger in parallel with the excitation rod might explain why the real-life notched stick is so efficient compared to  the laboratory-built device. This effect needs to be studied in greater detail in future experiments, in order to quantify its importance in the operation of the real-life toy.


%

\section{Conclusion}

Despite the simplicity of its design, the underlying physics of the notched stick is surprisingly rich. 
Here we focused on the causes of the motion of the notched stick's tip. Our main conclusions are the following. First, we demonstrated that the control of the direction of rotation is made possible by an anisotropic restoring force acting on the tip of the notched stick, caused by applying a perturbation force at an angle of 45\textdegree\,  with respect to the driving force. Remarkably, the direction of the rotation only depends on the angle of the perturbation force, and in particular is independent from the driving frequency, the temporal shape of the driving force, or the value of the damping factor. Secondly, we revealed a rich frequency behavior that allowed us to validate our model and measure its parameters with an experiment. We also found a strong dependence of the shape of the notched stick's trajectory on the value of the damping factor. More precisely, a significant energy loss in the system is necessary in order for the device to work outside of laboratory conditions, where the driving frequency cannot be controlled precisely. Finally, we identified an additional effect caused by the vertical oscillation of the perturbation finger, which could play a significant role in the operation of the notched stick.

The behavior of the real-life notched stick is a complex problem, and the rotation of the stick's tip probably results from several effects combined. Two mechanisms are identified in this paper. The second mechanism was only studied numerically in Sec.\,IV, and it would be of great interest to conduct dedicated experiments on this aspect. Other subtle effects would be worth tackling, such that the role of the shape of the cross-section of the notched stick. Indeed, a rectangular cross-section seems to induce a rotation without the need for a lateral perturbation force \cite{Leonard1937, Welch1973}.

Since it is able to convert linear vibration into a controllable rotational motion, the notched stick is a type of \textit{vibrot} \cite{Broseghini2019, Altshuler2013, Scholz2016}. Applications of such systems have been identified in engineering, in the context of screw or bolts loosening or tightening \cite{Petri2001}, or in robotics, for instance to build ultrasound-driven motors \cite{Liu2010}, or for the study of collective behaviors \cite{Scholz2018}. Similar vibration-induced rotations have been observed in various other systems with broken symmetry \cite{Nanda2016, Kondo2017, Bhadra2021, Peraza2019}.



Finally, armed with our new understanding, we can draw an analogy between the mechanism of the notched stick and birefringent materials in optics. In classical electromagnetism, it is common to assume that the negative electric charges in the matter are elastically bound to atomic nuclei. Birefringent materials exhibit an anisotropic electrical susceptibility, which corresponds to an anisotropic restoring force acting on the charge \cite{Feynman1964c}. Thus, the electric charge is analogous to our point mass model of the notched stick’s tip. In this analogy, the driving force corresponds to the Lorentz's force exerted on the charge by an incident linearly polarized light field. Note that since the oscillating electric charge radiates an electromagnetic wave, it must also lose kinetic energy, i.e. there is a damping force. In this picture, the change of polarization state caused by the propagation of light in the medium originates in the elliptical motion of the electric charges, which is similar to the elliptical motion's of the notched stick's tip. With this analogy, we explain why the notched stick has the same symmetries as waveplates. Indeed, rotating a quarter-waveplate by $\pi/2$ in front of an incident linearly polarized beam of light reverses the helicity of the output field.

\begin{acknowledgements}
The authors thank François Nadal, Laurent Potier and Robert Bédoret for helpful discussions.
\end{acknowledgements}

\section*{AUTHOR DECLARATIONS}
The authors have no conflicts to disclose.

\bibliography{bib.bib}

\begin{thebibliography}{32}%
\makeatletter
\providecommand \@ifxundefined [1]{%
 \@ifx{#1\undefined}
}%
\providecommand \@ifnum [1]{%
 \ifnum #1\expandafter \@firstoftwo
 \else \expandafter \@secondoftwo
 \fi
}%
\providecommand \@ifx [1]{%
 \ifx #1\expandafter \@firstoftwo
 \else \expandafter \@secondoftwo
 \fi
}%
\providecommand \natexlab [1]{#1}%
\providecommand \enquote  [1]{``#1''}%
\providecommand \bibnamefont  [1]{#1}%
\providecommand \bibfnamefont [1]{#1}%
\providecommand \citenamefont [1]{#1}%
\providecommand \href@noop [0]{\@secondoftwo}%
\providecommand \href [0]{\begingroup \@sanitize@url \@href}%
\providecommand \@href[1]{\@@startlink{#1}\@@href}%
\providecommand \@@href[1]{\endgroup#1\@@endlink}%
\providecommand \@sanitize@url [0]{\catcode `\\12\catcode `\$12\catcode `\&12\catcode `\#12\catcode `\^12\catcode `\_12\catcode `\%12\relax}%
\providecommand \@@startlink[1]{}%
\providecommand \@@endlink[0]{}%
\providecommand \url  [0]{\begingroup\@sanitize@url \@url }%
\providecommand \@url [1]{\endgroup\@href {#1}{\urlprefix }}%
\providecommand \urlprefix  [0]{URL }%
\providecommand \Eprint [0]{\href }%
\providecommand \doibase [0]{https://doi.org/}%
\providecommand \selectlanguage [0]{\@gobble}%
\providecommand \bibinfo  [0]{\@secondoftwo}%
\providecommand \bibfield  [0]{\@secondoftwo}%
\providecommand \translation [1]{[#1]}%
\providecommand \BibitemOpen [0]{}%
\providecommand \bibitemStop [0]{}%
\providecommand \bibitemNoStop [0]{.\EOS\space}%
\providecommand \EOS [0]{\spacefactor3000\relax}%
\providecommand \BibitemShut  [1]{\csname bibitem#1\endcsname}%
\let\auto@bib@innerbib\@empty
\bibitem [{\citenamefont {Scarnati}\ and\ \citenamefont {Tice}(1992)}]{Scarnati1992}%
  \BibitemOpen
  \bibfield  {author} {\bibinfo {author} {\bibfnamefont {J.~T.}\ \bibnamefont {Scarnati}}\ and\ \bibinfo {author} {\bibfnamefont {C.~J.}\ \bibnamefont {Tice}},\ }\bibfield  {title} {\bibinfo {title} {{The Hooey Machine}},\ }\href {https://doi.org/10.1080/00368121.1992.10113024} {\bibfield  {journal} {\bibinfo  {journal} {Science Activities}\ }\textbf {\bibinfo {volume} {29}},\ \bibinfo {pages} {30} (\bibinfo {year} {1992})}\BibitemShut {NoStop}%
\bibitem [{\citenamefont {Aubrecht}(1982)}]{Aubrecht1982}%
  \BibitemOpen
  \bibfield  {author} {\bibinfo {author} {\bibfnamefont {I.}~\bibnamefont {Aubrecht}, \bibfnamefont {Gordon~J.}},\ }\bibfield  {title} {\bibinfo {title} {{A mechanical toy: The gee‐haw whammy‐diddle}},\ }\href {https://doi.org/10.1119/1.2341165} {\bibfield  {journal} {\bibinfo  {journal} {The Physics Teacher}\ }\textbf {\bibinfo {volume} {20}},\ \bibinfo {pages} {614} (\bibinfo {year} {1982})}\BibitemShut {NoStop}%
\bibitem [{\citenamefont {{Schlichting}}\ and\ \citenamefont {{Backhaus}}(1988)}]{Schlichting1988}%
  \BibitemOpen
  \bibfield  {author} {\bibinfo {author} {\bibfnamefont {H.~J.}\ \bibnamefont {{Schlichting}}}\ and\ \bibinfo {author} {\bibfnamefont {U.}~\bibnamefont {{Backhaus}}},\ }\bibfield  {title} {\bibinfo {title} {{Zur Physik der Hui-Maschine}},\ }\href@noop {} {\bibfield  {journal} {\bibinfo  {journal} {Physik und Didaktik}\ }\textbf {\bibinfo {volume} {16,3}} (\bibinfo {year} {1988})}\BibitemShut {NoStop}%
\bibitem [{\citenamefont {{Satonobu}}\ \emph {et~al.}(1995)\citenamefont {{Satonobu}}, \citenamefont {{Ueha}},\ and\ \citenamefont {{Nakamura}}}]{Satonobu1995}%
  \BibitemOpen
  \bibfield  {author} {\bibinfo {author} {\bibfnamefont {J.}~\bibnamefont {{Satonobu}}}, \bibinfo {author} {\bibfnamefont {S.}~\bibnamefont {{Ueha}}},\ and\ \bibinfo {author} {\bibfnamefont {K.}~\bibnamefont {{Nakamura}}},\ }\bibfield  {title} {\bibinfo {title} {{A Study on the Mechanism of a Scientific Toy {\textquotedblleft}Girigiri-Garigari{\textquotedblright}}},\ }\href {https://doi.org/10.1143/JJAP.34.2745} {\bibfield  {journal} {\bibinfo  {journal} {Japanese Journal of Applied Physics}\ }\textbf {\bibinfo {volume} {34}},\ \bibinfo {pages} {2745} (\bibinfo {year} {1995})}\BibitemShut {NoStop}%
\bibitem [{\citenamefont {Bône}\ \emph {et~al.}(1994)\citenamefont {Bône}, \citenamefont {Morel},\ and\ \citenamefont {Boucher}}]{JCBone1994}%
  \BibitemOpen
  \bibfield  {author} {\bibinfo {author} {\bibfnamefont {J.-C.}\ \bibnamefont {Bône}}, \bibinfo {author} {\bibfnamefont {J.}~\bibnamefont {Morel}},\ and\ \bibinfo {author} {\bibfnamefont {M.}~\bibnamefont {Boucher}},\ }\href@noop {} {\emph {\bibinfo {title} {Mécanique générale}}}\ (\bibinfo  {publisher} {Dunod},\ \bibinfo {year} {1994})\BibitemShut {NoStop}%
\bibitem [{\citenamefont {Courty}\ and\ \citenamefont {Kierlik}(2004)}]{Courty2004}%
  \BibitemOpen
  \bibfield  {author} {\bibinfo {author} {\bibfnamefont {J.-M.}\ \bibnamefont {Courty}}\ and\ \bibinfo {author} {\bibfnamefont {E.}~\bibnamefont {Kierlik}},\ }\bibfield  {title} {\bibinfo {title} {Le bozo-bozo},\ }\href@noop {} {\bibfield  {journal} {\bibinfo  {journal} {Pour la Science}\ }\textbf {\bibinfo {volume} {318}} (\bibinfo {year} {2004})}\BibitemShut {NoStop}%
\bibitem [{\citenamefont {Welch}(1973)}]{Welch1973}%
  \BibitemOpen
  \bibfield  {author} {\bibinfo {author} {\bibfnamefont {S.~S.}\ \bibnamefont {Welch}},\ }\bibfield  {title} {\bibinfo {title} {{NOTES: What Makes it Turn?}},\ }\href {https://doi.org/10.1119/1.2350008} {\bibfield  {journal} {\bibinfo  {journal} {The Physics Teacher}\ }\textbf {\bibinfo {volume} {11}},\ \bibinfo {pages} {303} (\bibinfo {year} {1973})}\BibitemShut {NoStop}%
\bibitem [{\citenamefont {{Leonard}}(1937)}]{Leonard1937}%
  \BibitemOpen
  \bibfield  {author} {\bibinfo {author} {\bibfnamefont {R.~W.}\ \bibnamefont {{Leonard}}},\ }\bibfield  {title} {\bibinfo {title} {{An Interesting Demonstration of the Combination of Two Linear Harmonic Vibrations to Produce a Single Elliptical Vibration}},\ }\href {https://doi.org/10.1119/1.1991220} {\bibfield  {journal} {\bibinfo  {journal} {American Journal of Physics}\ }\textbf {\bibinfo {volume} {5}},\ \bibinfo {pages} {175} (\bibinfo {year} {1937})}\BibitemShut {NoStop}%
\bibitem [{\citenamefont {Wilson}(1998)}]{Wilson1998}%
  \BibitemOpen
  \bibfield  {author} {\bibinfo {author} {\bibfnamefont {J.~F.}\ \bibnamefont {Wilson}},\ }\bibfield  {title} {\bibinfo {title} {Parametric spin resonance for a spinner with an orbiting pivot},\ }\href {https://doi.org/https://doi.org/10.1016/S0020-7462(97)00020-6} {\bibfield  {journal} {\bibinfo  {journal} {International Journal of Non-Linear Mechanics}\ }\textbf {\bibinfo {volume} {33}},\ \bibinfo {pages} {189} (\bibinfo {year} {1998})}\BibitemShut {NoStop}%
\bibitem [{\citenamefont {{Marek}}\ \emph {et~al.}(2018)\citenamefont {{Marek}}, \citenamefont {{Badin}},\ and\ \citenamefont {{Plesch}}}]{Marek2018}%
  \BibitemOpen
  \bibfield  {author} {\bibinfo {author} {\bibfnamefont {M.}~\bibnamefont {{Marek}}}, \bibinfo {author} {\bibfnamefont {M.}~\bibnamefont {{Badin}}},\ and\ \bibinfo {author} {\bibfnamefont {M.}~\bibnamefont {{Plesch}}},\ }\bibfield  {title} {\bibinfo {title} {{Physics of the mechanical toy Gee-Haw Whammy Diddle}},\ }\href {https://doi.org/10.1038/s41598-018-22079-1} {\bibfield  {journal} {\bibinfo  {journal} {Scientific Reports}\ }\textbf {\bibinfo {volume} {8}},\ \bibinfo {eid} {3718} (\bibinfo {year} {2018})}\BibitemShut {NoStop}%
\bibitem [{\citenamefont {Broseghini}\ \emph {et~al.}(2019)\citenamefont {Broseghini}, \citenamefont {Ceccolini}, \citenamefont {Della~Volpe},\ and\ \citenamefont {Siboni}}]{Broseghini2019}%
  \BibitemOpen
  \bibfield  {author} {\bibinfo {author} {\bibfnamefont {M.}~\bibnamefont {Broseghini}}, \bibinfo {author} {\bibfnamefont {C.}~\bibnamefont {Ceccolini}}, \bibinfo {author} {\bibfnamefont {C.}~\bibnamefont {Della~Volpe}},\ and\ \bibinfo {author} {\bibfnamefont {S.}~\bibnamefont {Siboni}},\ }\bibfield  {title} {\bibinfo {title} {The notched stick, an ancient vibrot example},\ }\href {https://doi.org/10.1371/journal.pone.0218666} {\bibfield  {journal} {\bibinfo  {journal} {PLOS ONE}\ }\textbf {\bibinfo {volume} {14}},\ \bibinfo {pages} {1} (\bibinfo {year} {2019})}\BibitemShut {NoStop}%
\bibitem [{\citenamefont {{Caughey}}(1960)}]{Caughey1960}%
  \BibitemOpen
  \bibfield  {author} {\bibinfo {author} {\bibfnamefont {T.~K.}\ \bibnamefont {{Caughey}}},\ }\bibfield  {title} {\bibinfo {title} {{Hula-Hoop: An Example of Heteroparametric Excitation}},\ }\href {https://doi.org/10.1119/1.1935069} {\bibfield  {journal} {\bibinfo  {journal} {American Journal of Physics}\ }\textbf {\bibinfo {volume} {28}},\ \bibinfo {pages} {104} (\bibinfo {year} {1960})}\BibitemShut {NoStop}%
\bibitem [{\citenamefont {Cross}(2021)}]{Cross2021}%
  \BibitemOpen
  \bibfield  {author} {\bibinfo {author} {\bibfnamefont {R.}~\bibnamefont {Cross}},\ }\bibfield  {title} {\bibinfo {title} {Physics of a hula hoop},\ }\href {https://doi.org/10.1088/1361-6552/abd875} {\bibfield  {journal} {\bibinfo  {journal} {Physics Education}\ }\textbf {\bibinfo {volume} {56}},\ \bibinfo {pages} {025015} (\bibinfo {year} {2021})}\BibitemShut {NoStop}%
\bibitem [{\citenamefont {Bhattacharjee}(2013)}]{Bhattacharjee2013}%
  \BibitemOpen
  \bibfield  {author} {\bibinfo {author} {\bibfnamefont {S.}~\bibnamefont {Bhattacharjee}},\ }\href@noop {} {\bibinfo {title} {Synchronous motion in a devil's stick -- variation on a theme by kapitza}} (\bibinfo {year} {2013}),\ \Eprint {https://arxiv.org/abs/1307.6698} {arXiv:1307.6698} \BibitemShut {NoStop}%
\bibitem [{\citenamefont {Miller}(1955)}]{Miller1955}%
  \BibitemOpen
  \bibfield  {author} {\bibinfo {author} {\bibfnamefont {J.~S.}\ \bibnamefont {Miller}},\ }\bibfield  {title} {\bibinfo {title} {{The Notched Stick}},\ }\href {https://doi.org/10.1119/1.1933939} {\bibfield  {journal} {\bibinfo  {journal} {American Journal of Physics}\ }\textbf {\bibinfo {volume} {23}},\ \bibinfo {pages} {176} (\bibinfo {year} {1955})}\BibitemShut {NoStop}%
\bibitem [{\citenamefont {Laird}(1955)}]{Laird1955}%
  \BibitemOpen
  \bibfield  {author} {\bibinfo {author} {\bibfnamefont {E.~R.}\ \bibnamefont {Laird}},\ }\bibfield  {title} {\bibinfo {title} {{A Notched Stick}},\ }\href {https://doi.org/10.1119/1.1953047} {\bibfield  {journal} {\bibinfo  {journal} {American Journal of Physics}\ }\textbf {\bibinfo {volume} {23}},\ \bibinfo {pages} {472} (\bibinfo {year} {1955})}\BibitemShut {NoStop}%
\bibitem [{\citenamefont {Scott}(1956)}]{Scott1956}%
  \BibitemOpen
  \bibfield  {author} {\bibinfo {author} {\bibfnamefont {G.~D.}\ \bibnamefont {Scott}},\ }\bibfield  {title} {\bibinfo {title} {{Control of the Rotor on the Notched Stick}},\ }\href {https://doi.org/10.1119/1.1934275} {\bibfield  {journal} {\bibinfo  {journal} {American Journal of Physics}\ }\textbf {\bibinfo {volume} {24}},\ \bibinfo {pages} {464} (\bibinfo {year} {1956})}\BibitemShut {NoStop}%
\bibitem [{Vsa()}]{Vsauce}%
  \BibitemOpen
  \href@noop {} {\bibinfo {title} {The hui stick}},\ \bibinfo {howpublished} {\url{https://www.youtube.com/watch?v=lb7t1kI2x3o}}\BibitemShut {NoStop}%
\bibitem [{\citenamefont {Landau}\ and\ \citenamefont {Lifshitz}(1976)}]{Landau1976}%
  \BibitemOpen
  \bibfield  {author} {\bibinfo {author} {\bibfnamefont {L.~D.}\ \bibnamefont {Landau}}\ and\ \bibinfo {author} {\bibfnamefont {E.~M.}\ \bibnamefont {Lifshitz}},\ }\href@noop {} {\emph {\bibinfo {title} {Mechanics}}}\ (\bibinfo  {publisher} {Butterworth Heinemann},\ \bibinfo {year} {1976})\ Chap.~\bibinfo {chapter} {5}\BibitemShut {NoStop}%
\bibitem [{\citenamefont {Collett}(2005)}]{Collett2005}%
  \BibitemOpen
  \bibfield  {author} {\bibinfo {author} {\bibfnamefont {E.}~\bibnamefont {Collett}},\ }\href@noop {} {\emph {\bibinfo {title} {Field Guide to Polarization}}}\ (\bibinfo  {publisher} {SPIE Press, Bellingham, WA},\ \bibinfo {year} {2005})\BibitemShut {NoStop}%
\bibitem [{\citenamefont {{Feynman}}(1964{\natexlab{a}})}]{Feynman1964b}%
  \BibitemOpen
  \bibfield  {author} {\bibinfo {author} {\bibfnamefont {R.~P.}\ \bibnamefont {{Feynman}}},\ }\href@noop {} {\emph {\bibinfo {title} {{Feynman lectures on physics. Volume 1: Mainly Mechanics, Radiation and Heat}}}}\ (\bibinfo {year} {1964})\ Chap.~\bibinfo {chapter} {24}\BibitemShut {NoStop}%
\bibitem [{\citenamefont {{Feynman}}(1964{\natexlab{b}})}]{Feynman1964}%
  \BibitemOpen
  \bibfield  {author} {\bibinfo {author} {\bibfnamefont {R.~P.}\ \bibnamefont {{Feynman}}},\ }\href@noop {} {\emph {\bibinfo {title} {{Feynman lectures on physics. Volume 1: Mainly Mechanics, Radiation and Heat}}}}\ (\bibinfo {year} {1964})\ Chap.~\bibinfo {chapter} {23}\BibitemShut {NoStop}%
\bibitem [{\citenamefont {Altshuler}\ \emph {et~al.}(2013)\citenamefont {Altshuler}, \citenamefont {Pastor}, \citenamefont {Garcimartín}, \citenamefont {Zuriguel},\ and\ \citenamefont {Maza}}]{Altshuler2013}%
  \BibitemOpen
  \bibfield  {author} {\bibinfo {author} {\bibfnamefont {E.}~\bibnamefont {Altshuler}}, \bibinfo {author} {\bibfnamefont {J.~M.}\ \bibnamefont {Pastor}}, \bibinfo {author} {\bibfnamefont {A.}~\bibnamefont {Garcimartín}}, \bibinfo {author} {\bibfnamefont {I.}~\bibnamefont {Zuriguel}},\ and\ \bibinfo {author} {\bibfnamefont {D.}~\bibnamefont {Maza}},\ }\bibfield  {title} {\bibinfo {title} {Vibrot, a simple device for the conversion of vibration into rotation mediated by friction: Preliminary evaluation},\ }\href {https://doi.org/10.1371/journal.pone.0067838} {\bibfield  {journal} {\bibinfo  {journal} {PLOS ONE}\ }\textbf {\bibinfo {volume} {8}},\ \bibinfo {pages} {1} (\bibinfo {year} {2013})}\BibitemShut {NoStop}%
\bibitem [{\citenamefont {Scholz}\ \emph {et~al.}(2016)\citenamefont {Scholz}, \citenamefont {D’Silva},\ and\ \citenamefont {Pöschel}}]{Scholz2016}%
  \BibitemOpen
  \bibfield  {author} {\bibinfo {author} {\bibfnamefont {C.}~\bibnamefont {Scholz}}, \bibinfo {author} {\bibfnamefont {S.}~\bibnamefont {D’Silva}},\ and\ \bibinfo {author} {\bibfnamefont {T.}~\bibnamefont {Pöschel}},\ }\bibfield  {title} {\bibinfo {title} {Ratcheting and tumbling motion of vibrots},\ }\href {https://doi.org/10.1088/1367-2630/18/12/123001} {\bibfield  {journal} {\bibinfo  {journal} {New Journal of Physics}\ }\textbf {\bibinfo {volume} {18}},\ \bibinfo {pages} {123001} (\bibinfo {year} {2016})}\BibitemShut {NoStop}%
\bibitem [{\citenamefont {Petri}(2001)}]{Petri2001}%
  \BibitemOpen
  \bibfield  {author} {\bibinfo {author} {\bibfnamefont {P.}~\bibnamefont {Petri}},\ }\href@noop {} {\emph {\bibinfo {title} {Vibration-induced rotation}}}\ (\bibinfo  {publisher} {Massachusetts Institute of Technology},\ \bibinfo {year} {2001})\BibitemShut {NoStop}%
\bibitem [{\citenamefont {Liu}\ \emph {et~al.}(2010)\citenamefont {Liu}, \citenamefont {Chen}, \citenamefont {Liu},\ and\ \citenamefont {Shi}}]{Liu2010}%
  \BibitemOpen
  \bibfield  {author} {\bibinfo {author} {\bibfnamefont {Y.}~\bibnamefont {Liu}}, \bibinfo {author} {\bibfnamefont {W.}~\bibnamefont {Chen}}, \bibinfo {author} {\bibfnamefont {J.}~\bibnamefont {Liu}},\ and\ \bibinfo {author} {\bibfnamefont {S.}~\bibnamefont {Shi}},\ }\bibfield  {title} {\bibinfo {title} {Actuating mechanism and design of a cylindrical traveling wave ultrasonic motor using cantilever type composite transducer},\ }\href {https://doi.org/10.1371/journal.pone.0010020} {\bibfield  {journal} {\bibinfo  {journal} {PLOS ONE}\ }\textbf {\bibinfo {volume} {5}},\ \bibinfo {pages} {1} (\bibinfo {year} {2010})}\BibitemShut {NoStop}%
\bibitem [{\citenamefont {{Scholz}}\ \emph {et~al.}(2018)\citenamefont {{Scholz}}, \citenamefont {{Engel}},\ and\ \citenamefont {{P{\"o}schel}}}]{Scholz2018}%
  \BibitemOpen
  \bibfield  {author} {\bibinfo {author} {\bibfnamefont {C.}~\bibnamefont {{Scholz}}}, \bibinfo {author} {\bibfnamefont {M.}~\bibnamefont {{Engel}}},\ and\ \bibinfo {author} {\bibfnamefont {T.}~\bibnamefont {{P{\"o}schel}}},\ }\bibfield  {title} {\bibinfo {title} {{Rotating robots move collectively and self-organize}},\ }\href {https://doi.org/10.1038/s41467-018-03154-7} {\bibfield  {journal} {\bibinfo  {journal} {Nature Communications}\ }\textbf {\bibinfo {volume} {9}},\ \bibinfo {eid} {931} (\bibinfo {year} {2018})}\BibitemShut {NoStop}%
\bibitem [{\citenamefont {Nanda}\ \emph {et~al.}(2016)\citenamefont {Nanda}, \citenamefont {Singla},\ and\ \citenamefont {Karami}}]{Nanda2016}%
  \BibitemOpen
  \bibfield  {author} {\bibinfo {author} {\bibfnamefont {A.}~\bibnamefont {Nanda}}, \bibinfo {author} {\bibfnamefont {P.}~\bibnamefont {Singla}},\ and\ \bibinfo {author} {\bibfnamefont {M.~A.}\ \bibnamefont {Karami}},\ }\bibfield  {title} {\bibinfo {title} {Energy harvesting using rattleback: Theoretical analysis and simulations of spin resonance},\ }\href {https://doi.org/https://doi.org/10.1016/j.jsv.2015.12.032} {\bibfield  {journal} {\bibinfo  {journal} {Journal of Sound and Vibration}\ }\textbf {\bibinfo {volume} {369}},\ \bibinfo {pages} {195} (\bibinfo {year} {2016})}\BibitemShut {NoStop}%
\bibitem [{\citenamefont {Kondo}\ and\ \citenamefont {Nakanishi}(2017)}]{Kondo2017}%
  \BibitemOpen
  \bibfield  {author} {\bibinfo {author} {\bibfnamefont {Y.}~\bibnamefont {Kondo}}\ and\ \bibinfo {author} {\bibfnamefont {H.}~\bibnamefont {Nakanishi}},\ }\bibfield  {title} {\bibinfo {title} {Rattleback dynamics and its reversal time of rotation},\ }\href {https://doi.org/10.1103/PhysRevE.95.062207} {\bibfield  {journal} {\bibinfo  {journal} {Phys. Rev. E}\ }\textbf {\bibinfo {volume} {95}},\ \bibinfo {pages} {062207} (\bibinfo {year} {2017})}\BibitemShut {NoStop}%
\bibitem [{\citenamefont {Bhadra}\ \emph {et~al.}(2021)\citenamefont {Bhadra}, \citenamefont {Ghosh},\ and\ \citenamefont {Gupta}}]{Bhadra2021}%
  \BibitemOpen
  \bibfield  {author} {\bibinfo {author} {\bibfnamefont {S.}~\bibnamefont {Bhadra}}, \bibinfo {author} {\bibfnamefont {S.}~\bibnamefont {Ghosh}},\ and\ \bibinfo {author} {\bibfnamefont {S.}~\bibnamefont {Gupta}},\ }\bibfield  {title} {\bibinfo {title} {Emergent chirality and current generation},\ }\href {https://doi.org/10.1103/PhysRevResearch.3.043179} {\bibfield  {journal} {\bibinfo  {journal} {Phys. Rev. Res.}\ }\textbf {\bibinfo {volume} {3}},\ \bibinfo {pages} {043179} (\bibinfo {year} {2021})}\BibitemShut {NoStop}%
\bibitem [{\citenamefont {Peraza-Mues}\ and\ \citenamefont {Moukarzel}(2019)}]{Peraza2019}%
  \BibitemOpen
  \bibfield  {author} {\bibinfo {author} {\bibfnamefont {G.~G.}\ \bibnamefont {Peraza-Mues}}\ and\ \bibinfo {author} {\bibfnamefont {C.~F.}\ \bibnamefont {Moukarzel}},\ }\bibfield  {title} {\bibinfo {title} {Sustained rotation in a vibrated disk with asymmetric supports},\ }\href {https://doi.org/10.1088/1742-5468/ab417e} {\bibfield  {journal} {\bibinfo  {journal} {Journal of Statistical Mechanics: Theory and Experiment}\ }\textbf {\bibinfo {volume} {2019}},\ \bibinfo {pages} {103201} (\bibinfo {year} {2019})}\BibitemShut {NoStop}%
\bibitem [{\citenamefont {{Feynman}}(1964{\natexlab{c}})}]{Feynman1964c}%
  \BibitemOpen
  \bibfield  {author} {\bibinfo {author} {\bibfnamefont {R.~P.}\ \bibnamefont {{Feynman}}},\ }\href@noop {} {\emph {\bibinfo {title} {{Feynman lectures on physics. Volume 1: Mainly Mechanics, Radiation and Heat}}}}\ (\bibinfo {year} {1964})\ Chap.~\bibinfo {chapter} {33}\BibitemShut {NoStop}%
\end{thebibliography}%

\end{document}